\documentclass[12pt]{article}
\usepackage[latin1]{inputenc}
\usepackage{authblk}
\usepackage{amssymb}
\usepackage{amsmath}
\usepackage{amsfonts}
\usepackage{slashed}
\usepackage{graphicx}
\usepackage{setspace}
\usepackage{fancyhdr}
\usepackage{geometry}
\usepackage{plain}
\usepackage{multirow}
\usepackage{graphicx}
\usepackage{anysize}
\usepackage{placeins}
\usepackage{float}
\usepackage{color}
\usepackage{setspace}
\usepackage[backgroundcolor=green,linecolor=green]{todonotes}
\usepackage{array}
\usepackage[colorlinks=true]{hyperref}
\usepackage[all]{hypcap}
\usepackage{dcolumn}
\usepackage{array}
 \numberwithin{equation}{section}
 \hypersetup{colorlinks, citecolor=violet, linkcolor=violet,  urlcolor=violet}
 \marginsize{1cm}{1cm}{2cm}{2cm}

\begin{document}
 
\newcommand{\ket}[2]{\langle #1,#2\rangle}
\newcommand{\bra}[2]{\left[ #1,#2\right]}
\newcommand{\Pgen}[3]{\langle#1\lvert#2\lvert#3]}
\newcommand{\N}{$\mathcal{N}$}
\newcommand{\M}{$\mathcal{M}$}
\newcolumntype{d}{D{.}{.}{3}}

\makeatletter
\renewcommand*\env@matrix[1][c]{\hskip -\arraycolsep
  \let\@ifnextchar\new@ifnextchar
  \array{*\c@MaxMatrixCols #1}}
\makeatother

\thispagestyle{empty}

\begin{center}
\vspace{15mm}
\Large{\textbf{CSW-like Expansion for Einstein Gravity}} \\
\vspace{20mm}
\large\text{Brenda Penante$^{1,2}$, Sayeh Rajabi$^{1,2}$, Grigory Sizov$^{3}$}\\
\vspace{15mm}

\normalsize \textit{$^1$Perimeter Institute for Theoretical Physics, Waterloo, ON, N2L 2Y5, CA} \\
\vspace{2mm}
\vspace{2mm}
\normalsize\textit{$^2$Department of Physics and Astronomy $\&$ Guelph-Waterloo Physics Institute,} \\
\normalsize \textit{University of Waterloo, Waterloo, ON, N2L 3G1, CA}\\
\vspace{2mm}
\vspace{2mm}
\normalsize\textit{$^3$King's College London, Department of Mathematics, London, WC2R 2LS, UK}

\let\thefootnote\relax\footnotetext{b.penante@qmul.ac.uk, srajabi@perimeterinstitute.ca, grigory.sizov@kcl.ac.uk}
\end{center}
\vspace{20mm}

\abstract
Using the recent formula presented in \cite{He:2012er,Cachazo:2012pz} for the link representation of tree-level $\mathcal{N}=8$ supergravity amplitudes, we derived a CSW-like expansion for the Next-to-MHV 6- and 7-graviton amplitudes by using the global residue theorem; a technique introduced originally for Yang-Mills amplitudes \cite{ArkaniHamed:2009sx}. We analytically checked the equivalence of one of the CSW terms and its corresponding Risager's diagram \cite{Risager:2005vk}. For the remaining 6-graviton and all 7-graviton terms, we numerically checked the agreement with Risager's expansion. We showed that the conditions for the absence of contributions at infinity of the global residue theorem are satisfied for any number of particles. This means that our technique and Risager's should disagree starting at twelve particles where Risager's method is known to fail. 

\newpage

\section{Introduction and Main Results}

In their seminal work \cite{Cachazo:2004kj}, Cachazo, Svrcek and Witten (CSW) proposed an analytic expansion for tree-level Yang-Mills amplitudes. The rules of the expansion are similar to Feynman rules except for the vertices which are off-shell continuations of Maximally-Helicity-Violating (MHV) amplitudes. The off-shellness is provided by introducing an auxiliary spinor into the calculations. Risager \cite{Risager:2005vk}, inspired by the Britto-Cachazo-Feng-Witten (BCFW) recursive method \cite{Britto:2004ap,Britto:2005fq}, later presented a proof of the CSW rules to compute gluon amplitudes. In his approach, an auxiliary spinor similar to CSW's is used to apply a certain complex deformation on external momenta. Like in the BCFW method, the residue theorem is used to recover the non-deformed Yang-Mills amplitude. 

The simplicity of the CSW expressions in gauge theory motivated similar computations for gravity. However, when applied to gravity, Risager's method only produces graviton amplitudes up to eleven particles in the Next-to-MHV sector. The failure of Risager's method to write an MHV-expansion for NMHV gravity was first discovered by Bianchi et al.\ \cite{Bianchi:2008pu} using numerical techniques and later analytically confirmed by Benincasa et al.\ \cite{Benincasa:2007qj}. Also, recently the difference of the pure 12-graviton NMHV amplitude with its Risager's expansion was analytically calculated by Conde and one of the authors of this paper \cite{Conde:2012ik}. 

Here we briefly explain Risager's method. Throughout this paper, we only concentrate on pure (gluon or) graviton amplitudes in the NMHV sector which contain three negative helicity and $n-3$ positive helicity particles.
 Using the spinor-helicity notation for the on-shell momentum of particle $i$, $p^i=\lambda^i\tilde\lambda^i$, Risager's expansion is obtained from a complex deformation on the anti-holomorphic spinors, $\tilde\lambda^i$, of the external negative helicity particles. Since we restricted ourselves to the NMHV amplitudes, Risager's deformation with a complex
variable $z$ is given by
\begin{equation}
     \left\{ \begin{array}{rcc}
          \tilde\lambda^{(a)}(z)=\tilde\lambda^{(a)}+z\,\langle b\,c \rangle\tilde{\eta} \\
           \tilde\lambda^{(b)}(z)=\tilde\lambda^{(b)}+z\,\langle c\,a \rangle\tilde\eta \\
           \tilde\lambda^{(c)}(z)=\tilde\lambda^{(c)}+z\,\langle a\,b \rangle\tilde\eta
        \end{array} \right. \,      
         \label{eqn:def.Ris} 
\end{equation} 
where the negative helicities are labeled by $a$, $b$ and $c$, and $\tilde\eta$ is an arbitrary spinor. 

Through this deformation, the amplitude becomes a rational function of $z$, $\mathcal{M}_n(z)$. One can apply the residue theorem to construct the amplitude $\mathcal{M}_n$ from the poles of $\mathcal{M}_n(z)$. Risager's expansion, $\mathcal{M}_n^\text{Ris}$, is obtained by summing the residues of $\mathcal{M}_n(z)$: 

\begin{equation}
\mathcal{M}_{n}^{\textrm{Ris}}=\sum_{a,L^{+}}\mathcal{M}_{L}\left(\hat a^{-},L^{+},\left(-\hat{P}\right)^{-}\right)\frac{1}{(p_a+P_{L^+})^2}\,\mathcal{M}_{R}\left(\hat{P}^{+},\hat b^{-},\hat c^{-},R^{+}\right)\,,
\label{eqn:Risexp}
\end{equation}
where $\hat{P}=\hat{p}_a+P_{L^+}$ is Risager's deformed momentum flowing in the internal propagator, and $L^{+}$ ($R^{+}$) denotes the subset of external positive-helicity particles in the left (right) sub-amplitudes in \eqref{eqn:Risexp}. The hatted momenta are evaluated at the position of the poles which makes the corresponding propagator on-shell. Also, note that each term in the expansion is the product of two MHV amplitudes and a propagator.

As a consequence of the residue theorem, $\mathcal{M}_n(z)$ must vanish at infinity for the expansion to be valid, but the behavior for gravity is in fact $\mathcal{M}_n(z)\sim z^{n-12}$ when $z\to\infty$. This MHV-vertex expansion for gravity, hence, is not valid when applied to amplitudes of more than eleven gravitons. \\

As shown in \cite{Witten:GaugeAsStringInTwistor2003}, Yang-Mills MHV amplitudes are localized on lines in $\mathcal{Z}$ supertwistor space. The CSW expansion in Yang-Mills tells us that NMHV amplitudes are localized on pairs of lines in twistor space. Gravity MHV amplitudes, on the other hand, are not localized on lines because they have derivative of a $\delta$-function support on lines. However, in the NMHV sector one can write amplitudes as integrals over the Grassmannian $G(3,n)$ which is the space of $3$-planes in $\mathbb{C}^n$. A convenient way to represent a point in $G(3,n)$ is as a $3\times n$ matrix modulo a $GL(3)$ action. This matrix can be thought of as $n$ $3$-vectors which are the columns of the matrix. The columns are denoted by ${c}_a\in \mathbb{C}^3$, or equivalently $[c_a]\sim\mathbb{CP}^2$ as they are non-zero. We call the space of $3$-vectors the $C$-space and will discuss localizations in this space throughout this paper.

In Yang-Mills, NMHV amplitudes are also localized on pairs of lines in the $C$-space. Hence, localization in twistor space and $\mathbb{CP}^2$ are equivalent for Yang-Mills. We take CSW in gravity to mean that NMHV amplitudes are localized only on pairs of lines in the $C$-space. For this reason, we call the formulation in this paper a \emph{CSW-like} expansion for gravity.\\
 
Until recently, there has been an obvious lack of tools to explore a CSW-like expansion in gravity. One problem has been the absence of compact expressions for MHV amplitudes, the building blocks. This issue was overcome when Hodges presented an elegant formula for the MHV gravity amplitudes, analogous to the Parke-Taylor formula for Yang-Mills \cite{Hodges:2012ym}. Another obstacle has been the lack of a Grassmannian formulation. 

In \cite{Cachazo:2012kg}, Cachazo and Skinner proposed a manifestly permutation invariant formula to compute all tree-level $\mathcal{N}=8$ supergravity amplitudes in terms of higher degree rational maps to twistor space. The parity invariance and soft limits of the formula, two strong evidences for the proposal to be correct, were checked in \cite{Bullimore:2012cn}. This conjecture was later on proved in \cite{Cachazo:2012pz}, showing that the formula admits the BCFW recursion relations and also produces 3-particle MHV and $\overline{\text{MHV}}$ amplitudes. Moreover, the formula behaves at large momenta (large BCFW's complex variable, $z$) as tree-level gravity amplitudes do, $1/z^2$ \cite{ArkaniHamed:2008yf}. In \cite{He:2012er} and independently in the same work of Cachazo-Mason-Skinner \cite{Cachazo:2012pz} the formula was presented in the link representation. This formulation was indeed another missing ingredient to further explore CSW in gravity. Cachazo-Skinner formula and the link representation are reviewed in section \ref{sec:GravityFormula}.

In this note, we are looking for a formula analogous to the CSW one but for gravity. The analogy is not directly in twistor space but is in $C$-space. The main procedure is explained in section \ref{sec:CSW}.  We considered the link representation of the Cachazo-Skinner formula and applied a global residue theorem (GRT) to it \cite{griffiths2011principles}. The  residue theorem writes the amplitude as a sum over terms which coincide with different localizations of particles on a pair of lines in the $C$-space. These are in fact the CSW terms for NMHV gravity amplitudes. 

\begin{figure}[htb]
\centering
\label{fig:CSWgeneral}
\includegraphics[scale=0.4]{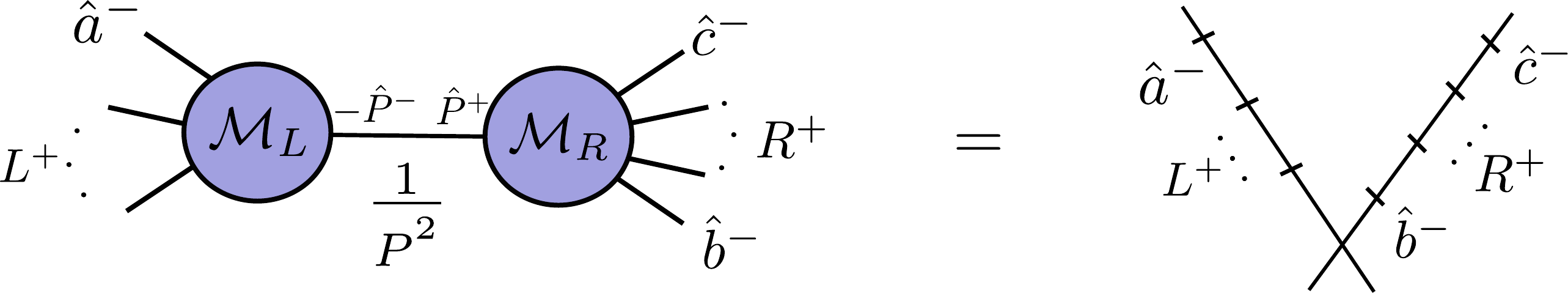}
\caption{The CSW-like localization of NMHV amplitudes on lines in $C$-space.}
\end{figure}

We computed the CSW terms for the NMHV 6- and 7-graviton scattering amplitudes in sections \ref{sec:6particles} and \ref{sec:7particles}, respectively.
The results obtained are compared to the known corresponding Risager's diagrams. We show that the two sides are in complete agreement. 
 Both theories, i.e., Yang-Mills and gravity, have an expansion in terms of two lines in $C$-space. Throughout this paper, we graphically represent each term in \eqref{eqn:Risexp} for gravity as two intersecting lines as shown in figure \ref{fig:CSWgeneral}.

For 6- and 7-graviton amplitudes, there are 21 and 45 Risager's diagrams respectively, which are the same as the localizations of the Cachazo-Skinner formula. This formula however, does not hold us back from exceeding 12-graviton amplitudes. Indeed, a power counting in section \ref{sec:Power Counting} shows that the global residue theorem is valid for any number of particles.   
In section \ref{sec:Conclusion} we discuss a possible reason for the discrepancy between Risager's and our technique. 

\section{Twistor String and Link Representation}
\label{sec:GravityFormula}

Recently, a new formula for the tree-level S-matrix of $ \mathcal{N}=8 $ supergravity for all R-charge sectors (labeled by $ d=k+1 $) was proposed by Cachazo and Skinner \cite{Cachazo:2012kg}:
\begin{equation}
\label{eq:g2n}
\mathcal{M}_n=\sum\limits_{d=0}^\infty\int\frac{\prod_{a=0}^d d^{4|8}\mathcal{Z}_a}{\text{Vol(GL(2))}}\det\hspace{0.1mm}'(\Phi)\det\hspace{0.1mm}'(\tilde{\Phi})\prod\limits_{i=1}^{n}d^2\sigma_i\, \delta^2(\lambda_i-\lambda(\sigma_1))\exp\left([\mu(\sigma_i)\,\tilde{\lambda}_i]+\chi^A\eta_A\right),
\end{equation}
whose ingredients are explained below:
\begin{itemize}
\item $ \mathcal{Z} $ is a holomorphic map from an $ n $-punctured Riemann sphere $ \Sigma $ with homogeneous coordinates $ (\sigma^{\underline{1}},\sigma^{\underline{2}}) $ to $ \mathcal{N}=8 $ supertwistor space $ \mathbb{CP}^{3|8} $ given by degree $ d $ polynomials
\begin{equation}
\mathcal{Z}^I(\sigma)=\sum\limits_{a=0}^d\mathcal{Z}^I_a(\sigma^{\underline{1}})^a(\sigma^{\underline{2}})^{d-a},
\end{equation}
\item $ \Phi $ and $ \tilde{\Phi} $ are $ n\times n $ matrices defined as
\begin{align}
\label{eq:phi}
\begin{split}
\Phi_{ij}&=\frac{\langle i\,j\rangle}{(i\,j)},\qquad\text{for }i\neq j,\\
\Phi_{ii}&=-\sum\limits_{j\neq i}\left\{\Phi_{ij}\frac{\prod_{k\neq i}(i\,k)}{\prod_{l\neq j}(j\,k)}\prod\limits_{a=0}^{\tilde{d}}\frac{(j\,p_a)}{(i\,p_a)}\right\},\qquad\tilde{d}=n-d-2,
\end{split}
\end{align}
and
\begin{align}
\label{eq:phitilde}
\begin{split}
\tilde{\Phi}_{ij}&=\frac{[i\,j]}{(i\,j)}\qquad\text{for }i\neq j,\\
\tilde{\Phi}_{ii}&=-\sum_{j\neq i}\tilde{\Phi}_{ij}\prod\limits_{a=0}^d\frac{(j\,p_a)}{(i\,p_a)},
\end{split}
\end{align}
where $ p_a $ are reference points in $ \Sigma $.
As written above, $ \Phi $ has rank $ d $ while $ \tilde{\Phi} $ has rank $ \tilde{d}=n-d-2 $. In order to obtain the reduced matrices $ \Phi^{\text{red}} $ and $ \tilde{\Phi}^{\text{red}} $ with non-vanishing determinant, one must first delete $ n-d $ rows and columns of $ \Phi $ and $ d+2 $ rows and columns of $ \tilde{\Phi} $. After this, the two $ \det\hspace{0.1mm}' $ in \eqref{eq:g2n} are defined as:
\begin{equation}
\det\hspace{0.1mm}'(\Phi)=\frac{|\Phi^{\text{red}}|}{|r_1\ldots r_d||c_1\ldots c_d|},\qquad\det\hspace{0.1mm}'(\tilde\Phi)=\frac{|\tilde\Phi^{\text{red}}|}{|r_1\ldots r_{d+2}||c_1\ldots c_{d+2}|},
\end{equation}
where $ |r_1\ldots r_d| $ and $ |c_1\ldots c_d| $ are the Vandermonde determinants of the \emph{remaining} rows and columns for $ \Phi $ and $ |r_1\ldots r_{d+2}| $ and $ |c_1\ldots c_{d+2}| $ of the \emph{deleted} rows and columns for $ \tilde{\Phi} $:
\begin{align}
\Phi:&\qquad|c_1\ldots c_d|=\prod\limits_{\substack{i,j\in \{\text{remaining}\}\\i<j}}(i\,j),\\
\tilde\Phi:&\qquad|c_1\ldots c_{d+2}|=\prod\limits_{\substack{i,j\in \{\text{deleted}\}\\i<j}}(i\,j),
\end{align}
and the same for $ |r_1\ldots r_d|,|r_1\ldots r_{d+2}| $.
\end{itemize}
It is known that \eqref{eq:g2n} is completely localized by the $ \delta $-functions, being the analog of the twistor string formulation for $ \mathcal{N}=4 $ super Yang-Mills tree amplitudes proposed by Witten \cite{Witten:GaugeAsStringInTwistor2003} and later studied by Roiban, Spradlin and Volovich \cite{Roiban:2004yf}. However, the constraints involve solving polynomials of high degree and summing the contribution of all solutions.

To simplify this problem, He \cite{He:2012er}, Cachazo, Mason and Skinner \cite{Cachazo:2012pz} wrote the amplitude in the so-called \emph{link representation}, which is a ``gauge fixed'' Grassmannian $ G(k,n) $ formulation in which $k= d+1 $ of the external particles are taken to be supertwistor space $ \mathcal{Z} $ eigenstates while the remaining ones are taken to be dual supertwistor space $ \mathcal{W} $ eigenstates. For amplitudes having only gravitons as incoming particles, we choose the gauge such that the $k$ columns associated with the negative helicity gravitons form an identity matrix. We label the gauge fixed columns by $ r $ and the remaining ones by $ a $. The unfixed variables $ c_{ra} $ are called \emph{link variables}.

In summary, the wavefunctions of the particles labeled by $ r $ are localized in $ \mathcal{Z} $ space, while the wavefunctions of the particles labeled by $ a $ are localized in $ \mathcal{W} $ space. In the link representation, each $3\times3$ minor becomes linear in the link variables, and the price to pay is that the amplitude is now a contour integral in $ (d-1)(\tilde{d}-1) $ variables.

As presented in \cite{Cachazo:2012pz}, the final product for the N$^{d-1}$MHV tree amplitude in the link representation reads
\begin{align}
\label{eq:linkrep}
\begin{split}
\mathcal{M}_{n,d}\left(\mathcal{Z}_r,\mathcal{W}_a\right)&=\int\left[\prod\limits_{r,a}\frac{d c_{ra}}{c_{ra}}\right]\frac{D^{n-1\,n}_{1\,2}}{\prod_a c_{1a}c_{2a}\prod_r c_{rn-1}c_{rn}}\\
&\times\left[\prod\limits_{\substack{r\neq 1,2\\ a\neq n-1,n}}D^{n-1\,n}_{1\,2}c_{1a}c_{2a}c_{rn-1}c_{rn}\frac{1}{\mathcal{V}_{an-1n}^{12r}}\right]\\
&\times \phi^{(d)}\left(\frac{\langle a\,b \rangle}{H^{ab}_{12}}\right)\phi^{(\tilde{d})}\left(\frac{[r\,s]}{H^{n-1n}_{rs}}\right)\exp\left(i c_{ra}\mathcal{Z}^r\mathcal{W}^a\right),
\end{split}
\end{align}
where
\begin{align}
D_{rs}^{ab}=\begin{vmatrix}
c_{ra} & c_{rb}\\
c_{sa} & c_{sb}
\end{vmatrix},\qquad\qquad  &H_{rs}^{ab}=\frac{D_{rs}^{ab}}{c_{ra}c_{rb}c_{sa}c_{sb}},
\end{align}
\begin{equation}
\label{eq:veronese}
\mathcal{V}_{an-1n}^{12r}=(1\,2\,r)(r\,a\,n-1)(n-1\,n\,1)(2\,a\,n)-(2\,r\,a)(a\,n-1\,n)(n\,1\,2)(r\,n-1\,1).
\end{equation}
We call $\mathcal{V}_{an-1n}^{12r}$ the Veronese polynomial and the reason for the name will be explained in the next section. $ \phi^{(d)},\phi^{(\tilde{d})} $ are the determinants present in \eqref{eq:g2n} with the following particular choice of removed rows and columns: using cyclicity, the labels are chosen such that particle 1 has negative helicity and particle $n$ has positive helicity. Then,
\begin{itemize}
\item from \eqref{eq:phi}, all rows and columns $a$ \emph{plus} $r=1$ are removed,
\item from \eqref{eq:phitilde}, all rows and columns $r$ \emph{plus} $a=n$ are removed.
\end{itemize}
In other words, $ \phi^{(d)} $ is the determinant of a $ d\times d $ matrix with elements
\begin{align}
\begin{split}
\phi_{rs}&=\frac{\langle r\,s\rangle}{H_{rs}^{n-1n}}\qquad\text{for  } r\neq s \quad\text{  and  }\quad r,s\neq 1,\\
\phi_{rr}&=-\sum\limits_{s\neq r}\frac{\langle r\,s\rangle}{H_{rs}^{n-1n}},
\end{split}
\end{align}
where as before, $ r $ and $s $ run over the particles labeled by $ \mathcal{Z} $ leaving out 1, and
$ \phi^{(\tilde{d})} $ is the determinant of a $ \tilde{d}\times \tilde{d} $ matrix with elements
\begin{align}
\begin{split}
\phi_{ab}&=\frac{[a\,b]}{H_{12}^{ab}}\qquad\text{for  } a\neq b \quad\text{  and  }\quad a,b\neq n,\\
\phi_{aa}&=-\sum\limits_{b\neq a}\frac{[a\,b]}{H_{12}^{ab}},
\end{split}
\end{align}
where the index $a$ runs over the set of particles localized in $ \mathcal{W} $ space, except for $ n $. \\

\section{A CSW-like Expansion for Gravity}
\label{sec:CSW}

In \cite{ArkaniHamed:2009sx}, a derivation of the CSW expansion for the $\mathcal{N}=4$ super Yang-Mills NMHV amplitudes from the Grassmannian is presented using the ``relaxing $\delta$-fuctions'' procedure described in section 3 of \cite{ArkaniHamed:2009sx}. Here we apply the same procedure to gravity, but instead of working in momentum-twistor space, we start from the link representation formula \eqref{eq:linkrep} written in momentum space.

Let us here summarize the procedure. To start with, it is useful to rewrite the products in \eqref{eq:linkrep} as
\begin{align*}
\prod\limits_{\substack{r\neq 1,2\\ a\neq n-1,n}}D^{n-1\,n}_{1\,2}c_{1a}c_{2a}c_{rn-1}c_{rn}= \left(D^{n-1\,n}_{1\,2}\right)^{(d-1)(\tilde{d}-1)}\left(\prod_{a\neq n,n-1} c_{1a}c_{2a}\right)^{d-1}\left(\prod_{r\neq 1,2} c_{rn-1}c_{rn}\right)^{\tilde{d}-1},
\end{align*}
so that
\begin{align*}
&\frac{D^{n-1\,n}_{1\,2}}{\prod_a c_{1a}c_{2a}\prod_r c_{rn-1}c_{rn}}\left[\prod\limits_{\substack{r\neq 1,2\\ a\neq n-1,n}}D^{n-1\,n}_{1\,2}c_{1a}c_{2a}c_{rn-1}c_{rn}\right]\\
=&\frac{\left(D^{n-1\,n}_{1\,2}\right)^{(d-1)(\tilde{d}-1)+1}}{(c_{1n-1}c_{1n}c_{2n-1}c_{2n})^2}\left(\prod_{a\neq n,n-1} c_{1a}c_{2a}\right)^{d-1}\left(\prod_{r\neq 1,2} c_{rn-1}c_{rn}\right)^{\tilde{d}-1}.
\end{align*}
Then, we can express \eqref{eq:linkrep} as
\begin{align}
\label{eq:GravityLink}
\begin{split}
\mathcal{M}_{n,d}=\int\prod\limits_{r,a}\frac{dc_{ra}}{c_{ra}}\prod\limits_{a\neq n, n-1}(c_{1a}c_{2a})^{d-2}\prod\limits_{r\neq 1,2}(c_{rn-1}c_{rn})^{\tilde{d}-2}\frac{\left(D_{12}^{n-1n}\right)^{(d-1)(\tilde{d}-1)+1}}{(c_{1n-1}c_{1n}c_{2n-1}c_{2n})^2}\\
\times \phi^{(d)}\left(\frac{\langle r\,s\rangle}{H_{rs}^{n-1n}}\right)\phi^{(\tilde{d})}\left(\frac{[a\,b]}{H_{12}^{ab}}\right)\prod\limits_{\substack{r\neq 1,2 \\ a\neq n,n-1}}\frac{1}{\mathcal{V}^{123}_{n-1na}}\exp\left(i c_{ra}\mathcal{W}^r\mathcal{Z}^a\right).
\end{split}
\end{align}
For the NMHV sector, we have $d=2$ and $\tilde{d}=n-4$; so the number of integration variables is $(d-1)(\tilde{d}-1)=n-5$. Also, we choose the gauge
\begin{equation}
\label{cmatrix}
\begin{pmatrix}
1 & 0 & 0 & c_{14} & c_{15} & \ldots & c_{1n} \\
0 & 1 & 0 & c_{24} & c_{25} & \ldots & c_{2n} \\
0 & 0 & 1 & c_{34} & c_{35} & \ldots & c_{3n} \\
\end{pmatrix}
\end{equation}
such that $r,s=1,2,3$ and $a,b=4,5,\ldots,n$. Then, \eqref{eq:GravityLink} written in momentum space reduces to
\begin{align}
\label{eq:GravityLinkNMHV}
\begin{split}
\mathcal{M}_{n,2}=\int\prod\limits_{r,a}\frac{dc_{ra}}{c_{ra}}(c_{3n-1}c_{3n})^{n-6}\frac{\left(D_{12}^{n-1n}\right)^{n-4}}{(c_{1n-1}c_{1n}c_{2n-1}c_{2n})^2}\phi^{(2)}\left(\frac{\langle r\,s\rangle}{H_{rs}^{n-1n}}\right)\phi^{(n-4)}\left(\frac{[a\,b]}{H_{12}^{ab}}\right)\\
\times \prod\limits_{a=4}^{n-2}\frac{1}{\mathcal{V}^{123}_{n-1na}}\prod\limits_{r=1}^3
\delta^2(\tilde{\lambda}_r+c_{ra}\tilde{\lambda}^a)\prod\limits_{a=4}^{n}\delta^2\left(\lambda_a-\lambda^r c_{ra}\right).
\end{split}
\end{align}
\newline
Now, starting from \eqref{eq:GravityLinkNMHV}, we perform the following steps: \\
\begin{enumerate}
\item Pull out momentum conservation 
\begin{equation}
\label{eq:momentumconservation} \prod\limits_{r=2}^3\delta^2(\tilde{\lambda}_r+c_{ra}\tilde{\lambda}^a)=\langle23\rangle^2\delta^4\left(\sum_{i=1}^n\lambda_i\tilde{\lambda}^i\right); \end{equation} 
\item Split the remaining $\tilde{\lambda}_1$ ``two component'' $\delta$-functions into two ``one-component'' $\delta$-functions by projecting it on two arbitrary linearly independent spinors $[X|$ and $[Y|$:
\[ \delta^2(\tilde{\lambda}_1+c_{1a}\tilde{\lambda}^a)=[XY]\delta([1X]+c_{1a}[aX])\delta([1Y]+c_{1a}[aY]);\]
\item Solve the system with all $\lambda$  $\delta$-functions together with $\delta([1Y]+c_{1a}[aY]) $,  and relax $ \delta([1X]+c_{1a}[aX]) $, that is to replace the $\delta$-function by one over its argument and treat the integral as a contour integral around the point where this argument is zero:
$$ \delta([1X]+c_{1a}[aX]) \longrightarrow\frac{1}{[1X]+c_{1a}[aX]}; $$
\item Since we relaxed one $\delta$-function, we increase by one the number of integration variables $t_i,\, i=1,2,\ldots,n-4$. The amplitude $\mathcal{M}_{n,2}$ is obtained by carrying the integrations on a contour defined around the zeros of the $n-5$ Veronese polynomials plus the zero of $[1X]+c_{1a}[aX]$. However, relaxing the $\delta$-function means treating this point as a pole, so we can use the global residue theorem to ``blow up the residue'' and write the amplitude as minus the sum of residues of the other poles of the integrand. Being more precise, suppose we perform $n-5$ integrations to solve all Veronese constraints. Then, we are left with one integration, say in the variable $t_1$, and the integrand is a function of $t_1$. To obtain the amplitude, we should compute the residue due to the only remaining constraint $[1X]+c_{1a}(t_1)[aX]=0$. However, if the integrand contains $m$ other poles for $t^*_{1,j},\, j=1,\ldots,m$, then 
\begin{equation}
\label{eq:cswexp}
\mathcal{M}_{n,2}=-\left.\sum\limits_{j=1}^m\text{Res}[\text{Integrand}]\right|_{t^*_{1,j}},
\end{equation}
where these poles correspond to $3\times3$ minors in the denominator. 
\end{enumerate}
These steps are the core of the CSW expansion for super Yang-Mills amplitudes, and in this work we apply the same procedure for the gravity formula \eqref{eq:GravityLinkNMHV} to calculate an analogous expansion for 6- and 7-graviton NMHV amplitudes.

Let us now explain how this expansion \eqref{eq:cswexp} coincides with CSW. We discuss the localization in the space of 3-vectors, $C$-space, which are the columns of a $3\times n$ matrix built from three $n$-vectors \eqref{cmatrix}. Both Yang-Mills and gravity amplitudes in the NMHV sector localize on pairs of lines in the $C$-space. For Yang-Mills, this localization is equivalent to the localization in twistor space, while for gravity the situation is different, as we know even the MHV amplitudes in gravity are not represented by lines in twistor space. 

Veronese map is a one from $G(2,n)$ to $G(k,n)$ which transforms world-sheet variables to $k$-vectors, columns of a $k\times n$ matrix. Veronese polynomial, defined in \eqref{eq:veronese}, when is set to zero, is a constraint for the columns of the $C$-matrix to be the same as columns of this $k\times n$ matrix.  

In the computations of residues of \eqref{eq:GravityLinkNMHV}, the $C$-matrix is evaluated on each of the poles of the integrand. This $3\times n$ matrix, after a rescaling of one of its rows, can be depicted on a 2-plane. Since we have $n-5$ Veronese polynomials in \eqref{eq:GravityLinkNMHV}, not all but some of the solutions make the points of the rescaled $C$-matrix lie on a conic. There also exist some non-conic configurations. The numerator, however, vanishes on the solutions leading to these configurations, so projects them out. Moreover, at any time, as there is a minor from the denominator set to zero, three of the points have to be collinear. In general, any five points define a unique conic. When three of these points are collinear, the conic is reducible and the configuration will be a degenerate conic which has three points on one line and the others spreading on any of the two lines. So, in summary, on the support of $n-5$ Veronese polynomials and the numerator we only see conic configurations. Each of the minors in the denominator then tells us which 3 points must be collinear. The outcomes are the terms in \eqref{eq:cswexp} which we define to be CSW terms. 

Here we refer to our expansion not as a ``CSW expansion for gravity'', but as a CSW-like expansion. The reason is that ``the canonical" way of generating the MHV expansion for gravity amplitudes is through Risager's procedure. It is not obvious a priori, however, that the procedure shown here will correspond to Risager's as happens for super Yang-Mills. That point being clear, we will refer to the channels as CSW-terms rather than CSW-like-terms for convenience of the notation.

\section{6-Graviton Computation}
\label{sec:6particles}
For the 6-graviton NMHV amplitude, we substitute $ n=6 $, $ d=\tilde{d}=2 $ in \eqref{eq:linkrep}. As already said, we choose a gauge fixing such that the indices $r,\,a$ take values $ r=1,2,3 $ and $ a=4,5,6 $. 
For computational purposes, here we write the amplitude not in twistor space but in momentum space, so the factor $\exp\left(i c_{ra}\mathcal{Z}^r\mathcal{W}^a\right)$ from \eqref{eq:linkrep} is Fourier transformed into a product of $\delta$-functions. Then \eqref{eq:linkrep} is written as

\begin{align}
\label{eq:6particles}
\begin{split}
\mathcal{M}(1^-,2^-,3^-,4^+,5^+,6^+)=\int\prod\limits_{\substack{a=1,2,3\\r=4,5,6}}\frac{dc_{ra}}{c_{ra}}(H_{12}^{56})^2
\phi^{(2)}\left(\frac{[a\,b]}{H^{ab}_{12}}\right)\phi^{(2)}\left(\frac{\langle r\,s\rangle}{H^{56}_{rs}}\right)\\
\times
\frac{1}{\mathcal{V}}\prod_{a=4}^6\delta^2(\lambda_a-c_{ra}\lambda^r)\prod_{r=1}^3\delta^2(\tilde\lambda_r+c_{ra}\tilde\lambda^a),
\end{split}
\end{align}
where
\begin{align}
H_{12}^{56}=\frac{c_{15}c_{26}-c_{16}c_{25}}{c_{15}c_{26}c_{16}c_{25}}\quad\text{ and }\quad \mathcal{V}\equiv\mathcal{V}^{456}_{123}.
\end{align}
The functions $\phi^{(2)}$ entering the formula \eqref{eq:6particles} are determinants of arbitrary $2\times 2$ minors of the following rank-2 matrices:
\begin{align}
\Phi'=\left(\begin{array}{ccc}
-\frac{\langle 12\rangle}{H_{12}^{56}}-\frac{\langle 13\rangle}{H_{13}^{56}} & \frac{\langle 12\rangle}{H_{12}^{56}} & \frac{\langle 13\rangle}{H_{13}^{56}} \\
\frac{\langle 12\rangle}{H_{12}^{56}} & -\frac{\langle 12\rangle}{H_{12}^{56}}- \frac{\langle 23\rangle}{H_{23}^{56}} & \frac{\langle 23\rangle}{H_{23}^{56}} \\
\frac{\langle 13\rangle}{H_{13}^{56}} & \frac{\langle 23\rangle}{H_{23}^{56}} & -\frac{\langle 13\rangle}{H_{13}^{56}} -\frac{\langle 23\rangle}{H_{23}^{56}} 
\end{array}\right),
\end{align}
\begin{align}
\Phi=\left(\begin{array}{ccc}
- \frac{[45]}{H_{12}^{45}}   -  \frac{[46]}{H_{12}^{46}} &  \frac{[45]}{H_{12}^{45}}   &  \frac{[46]}{H_{12}^{46}}  \\
\frac{[45]}{H_{12}^{45}} & -\frac{[45]}{H_{12}^{45}}- \frac{[56]}{H_{12}^{56}}  &  \frac{[56]}{H_{12}^{56}} \\
 \frac{[46]}{H_{12}^{46}} & \frac{[56]}{H_{12}^{56}}	 &  - \frac{[46]}{H_{12}^{46}} -\frac{[56]}{H_{12}^{56}}  \\
\end{array}\right).
\end{align}
There are 9 integration variables and $12$ bosonic $\delta$-functions, four of which provide momentum-conservation \eqref{eq:momentumconservation}.
After pulling out the momentum conservation the integral \eqref{eq:6particles} contains $8$ constraints and $9$ integration variables, so it should be complemented by the choice of an integration contour. The amplitude is given by integration along the contour encircling ${\cal V}=0$. 

Now let us start the transformations leading  from the integral representation \eqref{eq:6particles} to the CSW-like expansion of the gravity amplitude.
First, according to the procedure explained in section \ref{sec:CSW}, we split

\begin{align}
\begin{split}
\delta^2(\tilde\lambda_1+c_{14}\tilde\lambda_4+c_{15}\tilde\lambda_5+c_{16}\tilde\lambda_6)=[XY]
&\delta\left([1X]+c_{14}[4X]+c_{15}[5X]+c_{16}[6X]\right)\\
\times &\delta\left([1Y]+c_{14}[4Y]+c_{15}[5Y]+c_{16}[6Y]\right), \end{split}
\end{align}
and relax $\delta\left([1X]+c_{1a}[aX]\right)$. All the factors in the denominator can then be combined into two functions: 
\begin{align}
&f_1=c_{15}c_{16}c_{25}c_{26}c_{34}(156)(256)(345)(364)\left([1X]+ c_{1a}[aX]\right),\\
&f_2={\cal V}.
\end{align}
The integral is now in two complex variables and the amplitude is then given by a multi-dimensional residue
at the poles where both $[1X]+c_{1a}[aX]$ and ${\cal V}$ are zero.

\begin{figure}[htb]
\centering
\label{fig:CSWexample6}
\includegraphics[scale=0.5]{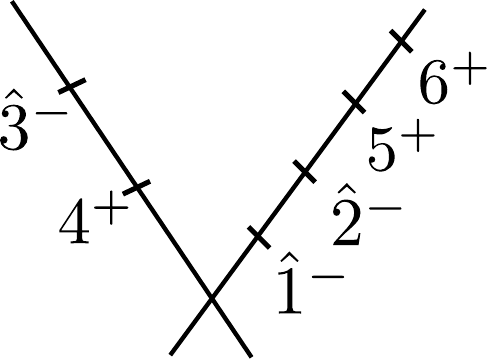}
\caption{A configuration associated with $\{(156),(126)\}$ and $\{(256),(126)\}$.}
\end{figure}

According to the global residue theorem, we can also write the amplitude as minus the sum of all other residues in which ${\cal V}$ and one of the factors in $f_1$ other than $[1X]+ c_{1a}[aX]$ are set to zero. Explicitly speaking, denoting a residue by the factors set to zero to compute it, then
\begin{align}
\begin{split}
\mathcal{M}_{6,2}=\{{\cal V},[1X]+ c_{1a}[aX]\}=-&\{{\cal V},c_{15}\}-\{{\cal V},c_{16}\}-\{{\cal V},c_{25}\}-\{{\cal V},c_{26}\}-\{{\cal V},c_{34}\}\\
-&\{{\cal V},(156)\}-\{{\cal V},(256)\}-\{{\cal V},(345)\}-\{{\cal V},(364)\}.
\end{split}
\end{align}
Notice that, since $\cal V$ is a polynomial of degree 4, each term splits yet into 4 terms. Notice also that many residues contribute to the same configuration, e.g., $\{(156),(126)\}$ and $\{(256),(126)\}$ shown in figure \ref{fig:CSWexample6}. The way to carefully deal with this and take care of all contributions for a given diagram will be explained in section \ref{sec:regulator}. \\
\newline

Carrying on the computation, the $\delta$-functions which we keep, impose the following set of seven equations
\begin{equation}
\begin{cases}
[1Y]+c_{14}[4Y]+c_{15}[5Y]+c_{16}[6Y]=0,\\
\lambda_4=c_{14}\lambda_1+c_{24}\lambda_2+c_{34}\lambda_3, \\
\lambda_5=c_{15}\lambda_1+c_{25}\lambda_2+c_{35}\lambda_3,\\
\lambda_6=c_{16}\lambda_1+c_{26}\lambda_2+c_{36}\lambda_3,\\
\end{cases}
\label{eq:system}
\end{equation}
or equivalently
\begin{equation}
\begin{cases}
[1Y]+c_{14}[4Y]+c_{15}[5Y]+c_{16}[6Y]=0,\\
c_{14}\langle 12\rangle+c_{34}\langle 32\rangle-\langle 42\rangle=0, \\
c_{24}\langle 21\rangle+c_{34}\langle 31\rangle-\langle 41\rangle=0,\\

c_{15}\langle 15\rangle+c_{25}\langle 25\rangle+c_{35}\langle 35\rangle=0, \\
c_{25}\langle 21\rangle+c_{35}\langle 31\rangle-\langle 51\rangle=0,\\

c_{16}\langle 16\rangle+c_{26}\langle 26\rangle+c_{36}\langle 36\rangle=0, \\
c_{26}\langle 21\rangle+c_{36}\langle 31\rangle-\langle 61\rangle=0.
\end{cases}
\label{eq:systemimproved}
\end{equation}
The last transformation is performed for computational convenience and multiplies the Jacobian by a factor of $\langle 21\rangle \langle 51\rangle \langle 61\rangle$. \\
\newline
Let us denote by $S$ the system of equations formed by \eqref{eq:systemimproved} and two additional equations: ${\cal V}$=0 and $(abc)=0$, where $(abc)$ is one of the minors in the denominator. Then, an individual residue is given by the following expression evaluated on the solution of the system $S$:
\begin{align}
\frac{1}{J}\frac{\left([45][56]c_{15}c_{25}(364)+\text{cyclic}(4,5,6)\right)\left(\langle 12\rangle\langle 23\rangle c_{25}c_{26}(256)+\text{cyclic}(1,2,3)\right)(abc)}{(135)(136)(235)(236)(124)(156)(256)(345)(364)}\frac{[XY]}{\left([1X]+ c_{1a}[aX]\right)},
\label{eq:almost}
\end{align}
where $J$ is the Jacobian of solving the system $S$. The precise way of calculating the residues will be explained later in section \ref{sec:regulator}.

\subsection{Analytical Computation of $\{(135),(246)\}$}

We calculate the residue of the integrand at the point $\{(135),(246)\}$. To see that this is indeed a simple multi-dimensional pole we
represent the Veronese polynomials in the form ${\cal V}=(123)(345)(561)(246)-(234)(456)(612)(351)$. Setting (135) to zero makes $f_1$ vanish. Setting additionally (246) to zero sets both $f_2$ and ${\cal V}$ to zero and, as one can check, does not add any zeros among the factors in the denominator.

Using Mathematica to find the solution of the system \eqref{eq:system}, substituting it into \eqref{eq:almost} and simplifying the result, we obtain
\begin{align}
\frac{[46]\langle 13\rangle^6\langle 2|4+6|5]\langle 2|4+6|Y]^5}{\langle 15\rangle \langle 24\rangle \langle 26\rangle \langle 35\rangle \langle 46\rangle\langle 1|3+5|Y] \langle 4|2+6|Y]\langle 6|2+4|Y] \langle 5|1+3|Y] \langle 3|1+5|Y](p_1+p_3+p_5)^2} .
\end{align}
One can see that this result coincides with a Risager's term which has the form
\begin{align}
\mathcal{M}_L(\hat{1}^-,\hat{3}^-,5^+,\hat{P}^+)\dfrac{1}{P^2} \mathcal{M}_R(\hat{2}^-,4^+,6^+,-\hat{P}^-),
\end{align}
where $P=p_2+p_4+p_6$ is the momentum flowing through the link which is set on-shell and $\hat P$ is the Risager-deformed momentum $\hat{P}=\hat{p}_2+p_4+p_6$ as in \eqref{eqn:Risexp}.

\subsection{Numerical Check of All Residues for 6 Gravitons: Taming Singular Configurations in Multi-Dimensional Residues}
\label{sec:regulator}

In order to numerically calculate the other channels, an immediate application of the global residue theorem leads to a computational problem: the part of the integrand besides the two zero factors in the denominator, which is supposed to be finite, can become ambiguous. This happens because the numerator and the denominator go to zero simultaneously and do not cancel with each other. 

\begin{figure}[h]
\centering
\label{fig:6example}
\includegraphics[scale=0.7]{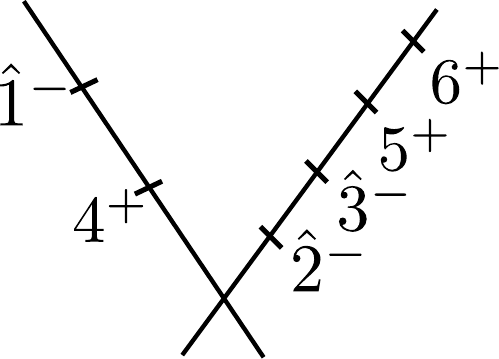}
\caption{An example of a 2-4 channel.}
\end{figure}

Consider for example the 2-4 channel shown in figure \ref{fig:6example} in which particles 1 and 4 belong to $\mathcal{M}_L$ and particles 2, 3, 5 and 6 belong to $\mathcal{M}_R$.
 \\
One way to generate this configuration is by setting $\mathcal{V}$ and $(256)$ to zero. Looking at the denominator of \eqref{eq:almost}, however, we see that $(235)$ and $(236)$ are ``accidentally'' set to zero as well. For the residue to be finite, the numerator of \eqref{eq:almost} should also go to zero at the solution. Restricting ourselves to the submanifold of $G(3,6)$ defined by $\mathcal{V}=0$ equation, we can write all minors that go to zero as a multiple of say (235). More precisely, in this example we can use $\mathcal{V}=(462)(235)(514)(631)-(623)(351)(146)(254)=0$\footnote{Keep in mind that the condition $\mathcal{V}=0$ is permutation invariant, so we can make a choice of ordering of particles which is the most convenient for our purposes.} to write $(236)\propto(235)$. Playing this game with all the minors that go to zero in the integrand, we note that both the numerator and denominator vanish as $(235)^5$, so they cancel and we can finally obtain a finite residue.

Although this method gives the correct results in this example, it requires an individual treatment of each residue, making the computations rather laborious. More importantly, it cannot necessarily be applied to higher-point computations. For this reason, we use a procedure which can be directly applied in all cases which consists of the use of a \emph{regulator}.

The singular residues are points in the Grassmannian on which the denominator of the integrand has a zero of second order or higher. The idea is to add to each minor in the denominator a small constant $\epsilon$, the regulator, 
\begin{equation}
\frac{1}{(abc)}\rightarrow \frac{1}{(abc)+\epsilon},
\end{equation}
which separates the higher order zeros in many zeros of first order. 
Also to make sure that the regulated minors do not accidentally vanish simultaneously, we may add different values of $\epsilon$ to different minors. After this, all poles are simple and the global residue theorem can be applied without restriction. Moreover, there is no need to regulate the Veronese polynomials, since they do not factorize when minors are regulated ($\epsilon\neq0$).

The Veronese polynomials and each of the regulated minors in the denominator make a system of equations set to zero. On each solution of each system of equations, there might be some minors which nearly vanish. We compute all possible minors on each solution and collect the almost-zero ones. These minors now define the localization of particles on the two crossing lines which makes a CSW channel. As an example, in figure \ref{fig:6example}, the only minors which are close to zero are $(235)$, $(236)$, $(256)$ and $(356)$. Hence, particles 2, 3, 5 and 6 lie on a line. And, there is also a line passing through any two points; 1 and 4 here. As it is also clear from this example, many solutions may contribute to a given localization. Here, residues of the integrand on the solutions of the following systems of equations contribute to this channel: $\mathcal{V}=(235)+\epsilon=0$, $\mathcal{V}=(236)+\epsilon=0$ and $\mathcal{V}=(256)+\epsilon=0$. In the end, all contributions must be summed to give the corresponding CSW term.

\begin{figure}[h]
\label{fig:Factorizations6}
\centering
\includegraphics[scale=0.5]{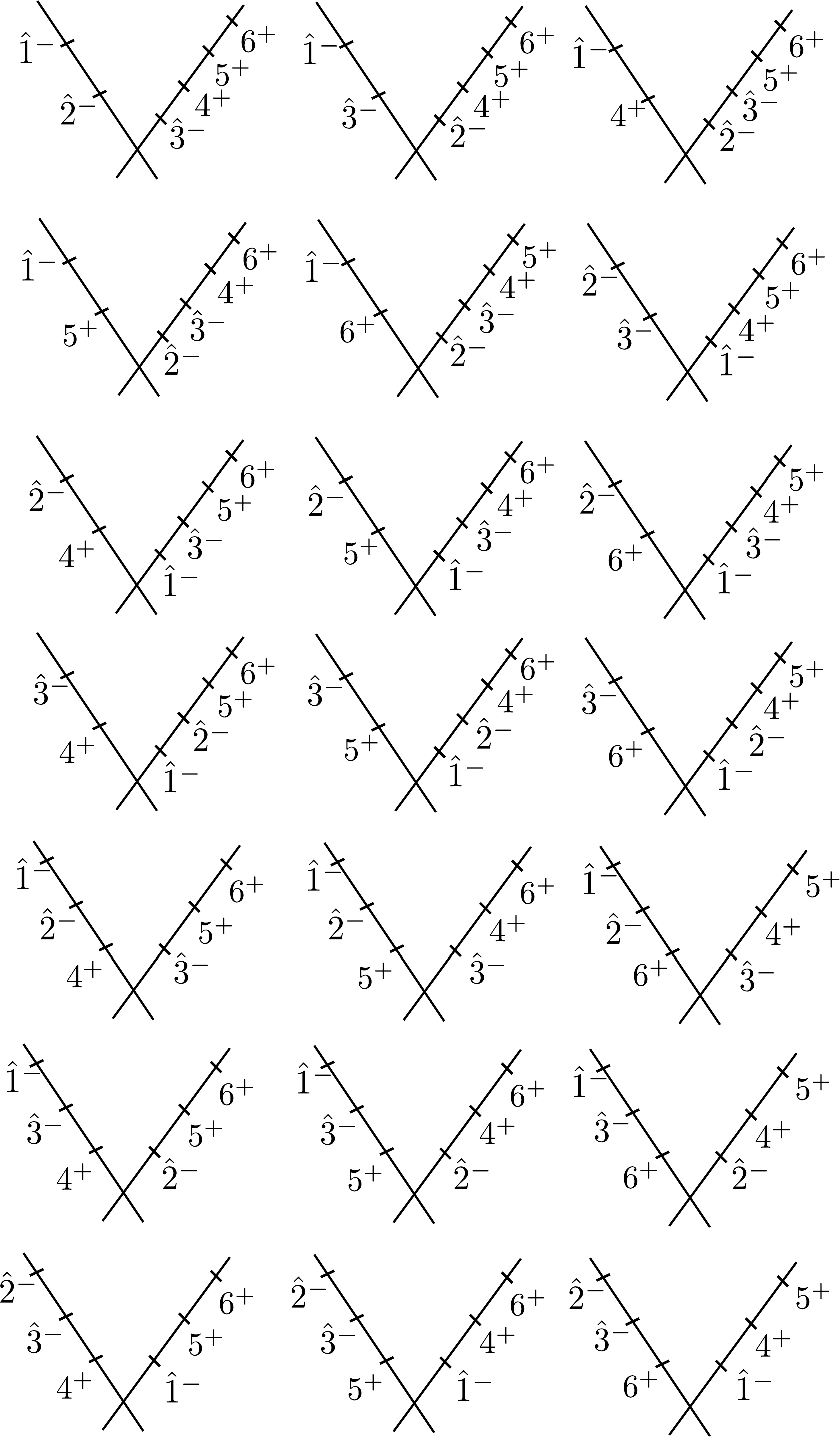}
\caption{6-graviton channels}
\end{figure}

This way, we computed the numerical values of the 21 configurations of 6 gravitons, depicted in figure \ref{fig:Factorizations6}.
The numerical values obtained are compared to the known corresponding Risager terms. As a result of this comparison we found, as happens to Yang-Mills, a complete agreement between our procedure and Risager's. Numerical results are given in the appendix for both 6- and 7-graviton cases.\\

\section{7-Graviton Computation}
\label{sec:7particles}

For $n=7$, $d=2$ and $\tilde d=3$, so the general formula \eqref{eq:linkrep} is
\begin{align}
\begin{split}
\mathcal{M}(1^-,2^-,3^-,4^+,5^+,6^+,7^+)=\int\prod\limits_{\substack{r=1,2,3\\a=4,..,7}}\frac{dc_{ra}}{c_{ra}}c_{36}c_{37}\frac{(367)^3}{(c_{16}c_{17}c_{26}c_{27})^2}\\
\times
\phi^{(3)}\left(\frac{[a\,b]}{H_{12}^{ab}}\right)\phi^{(2)}\left(\frac{\left\langle r\,s\right\rangle}{H_{rs}^{67}}\right)\frac{1}{\mathcal{V}^{123}_{467}\mathcal{V}^{123}_{567}}\prod_r\delta^2(\tilde\lambda_r+c_{ra}\tilde\lambda^a)\prod_a\delta^2(\lambda_a-c_{ra}\lambda^r).
\end{split}
\end{align}
Using the definitions of $\phi$, one can rewrite the integrand as a rational expression whose denominator is
\begin{align}
\label{eq:den7}
(136)(137)(236)(237)(124)(125)(167)(267)(345)(346)(347)(356)(357)\mathcal{V}^{123}_{467}\mathcal{V}^{123}_{567}.
\end{align}
According to the general procedure of section \ref{sec:CSW}, we relax one component of the $\tilde{\lambda}_1$ $\delta$-functions. After this, we get an additional factor of $[1X]+\sum_a c_{1a}[aX]$ in the denominator.
To apply a global residue theorem these factors should be split into three functions:
\begin{align}
\label{eq:f123}
\begin{split}
&f_1=(136)(137)(236)(237)(124)(125)(167)(267)(345)(346)(347)(356)(357)\left([1X]+\sum\limits_a c_{1a}[aX]\right),\\
&f_2=\mathcal{V}^{123}_{467},\\
&f_3=\mathcal{V}^{123}_{567}.
\end{split}
\end{align}
The amplitude is the residue of the integrand at the point where $f_2$, $f_3$ and the last factor in $f_1$ go to zero. 
Like in the 6-particle case, we get different CSW terms by computing residues at points where $f_2,f_3$ and one of the minors in $f_1$ go to zero.
The GRT then tells us that all the CSW terms sum to the full amplitude, as it naturally should be:

\begin{align}
\mathcal{M}_{7,2}=\left\{\left([1X]+\sum\limits_a c_{1a}[aX]\right),f_2,f_3\right\}=-\{(136),f_2,f_3\}-\{(137),f_2,f_3\}-\ldots-\{(357),f_2,f_3\}.
\end{align}
Notice that every term in the right-hand side can in principle contain more than one CSW term, because the system 
\begin{align}
\begin{cases}
\text{minor}=0,\\
 \mathcal{V}^{123}_{467}=0,\\
 \mathcal{V}^{123}_{567}=0.
 \end{cases}
 \label{eq:minorsystem}
\end{align}
 can have more than one solution.

To compute the residues we perform a procedure similar to the one described in section \ref{sec:regulator}. First we introduce the regulator to every minor in \eqref{eq:den7} and write a system of linear equations composed of the $\delta$-functions and three equations setting a minor and the two $\cal V$'s to zero. After this, all residues are free from any accidental ambiguity and out of the $13\times 4\times 4$ solutions, many of them sum to give the 45 possible configurations.

If the configuration is such that these two Veronese polynomials are equal to zero, but some other Veronese polynomials which can be made of particles $1,\ldots,7$ are not zero, then this configuration does not contribute, since the corresponding residue is zero (such configurations are called {\it spurious solutions}). 

\section{The General $n$ NMHV Case: Power Counting}
\label{sec:Power Counting}
All the reasoning presented in section \ref{sec:CSW} relies on the assumption that the global residue theorem is applicable and our aim now is to verify when this is the case for gravity. To do so, we analyse how the integrand behaves for large values of the integration variables, that is, we look for $n$ such that there is no pole at infinity. 

In order to perform the power counting, consider formula \eqref{eq:GravityLinkNMHV} which we repeat here for convenience:
\begin{align*}
\begin{split}
\mathcal{M}_{n,2}=\int\prod\limits_{r,a}\frac{dc_{ra}}{c_{ra}}\prod\limits_{a=4}^{n-2}(c_{3n-1}c_{3n})^{n-6}\frac{\left(D_{12}^{n-1n}\right)^{n-4}}{(c_{1n-1}c_{1n}c_{2n-1}c_{2n})^2}\phi^{(2)}\left(\frac{\langle r\,s\rangle}{H_{rs}^{n-1n}}\right)\phi^{(n-4)}\left(\frac{[a\,b]}{H_{12}^{ab}}\right)\\
\times \prod\limits_{a=4}^{n-2}\frac{1}{\mathcal{V}^{123}_{n-1na}}\prod\limits_{r=1}^3
\delta^2(\tilde{\lambda}_r+c_{ra}\tilde{\lambda}^a)\prod\limits_{a=4}^{n}\delta^2\left(\lambda_a-\lambda^r c_{ra}\right).
\end{split}
\end{align*}
We know from section 2.4 of \cite{ArkaniCachazo:DualityS-matrix2009}, that each minor is at most of degree one. We also know that in the link representation we can write each link variable and the factors $ D_{12}^{n-1n} $ as minors, so they are also linear in $t_i$, and keep in mind that each Veronese polynomial is of degree 4 in $t_i$. \\
\newline
Then, let us consider one variable $t$ and look at how each factor in the integrand of \eqref{eq:GravityLinkNMHV} scales with it:
\begin{align*}
&\prod\limits_{r,a}\frac{1}{c_{ra}}\sim  t^{9-3n},\\
&(c_{3n-1}c_{3n})^{n-6}\sim t^{2n-12},\\
&\frac{\left(D_{12}^{n-1n}\right)^{n-4}}{(c_{1n-1}c_{1n}c_{2n-1}c_{2n})^2}\sim \frac{t^{n-4}}{t^8}=t^{n-12},\\
&\phi^{(n-4)}\left(\frac{[ab]}{H_{12}^{ab}}\right)\sim t^{3(n-4)},\\
&\phi^{(2)}\left(\frac{\langle rs\rangle}{H_{rs}^{n-1n}}\right)\sim t^6,\\
&\prod\limits_{a\neq n,n-1}\delta\left(\mathcal{V}^{123}_{n-1na}\right)\sim t^{-4(n-5)},\\
&\text{Relaxed $\delta$-function}\sim t^{-1}.
\end{align*}
Summing up, we get that the integrand scales as $t^{-n-2}$. 
The condition for the global residue theorem to apply is that the integrand should decay faster than $t^{-(n-4)}$ for each integration variable, that is $-n-2 \leqslant -n+4$. This condition is satisfied $\forall\, n$, so we conclude that the CSW-like expansion will be valid for all NMHV graviton amplitudes.

\section{Conclusions}
\label{sec:Conclusion}
Building up on the recent results for $\mathcal{N}=8$ supergravity amplitudes~--- Hodge's MHV formula, the Cachazo-Skinner formula and its link representation~--- we carried on the appliance of the well-known techniques used for super Yang-Mills amplitudes to explore the new gravity territory.

From the link representation, we derived a CSW-like expansion for pure 6- and 7-graviton NMHV amplitudes. It was not obvious, however, that this expansion would coincide with Risager's, the known way up to now to produce an MHV-vertex expansion. We observed a complete agreement between the two expansions in the cases we worked on.

The most exciting feature of the new expansion is that in principle it works for any $n$ for $d=2$, while Risager's method fails for $n\geq 12$. Further studies are necessary to verify if and when the two expansions stop agreeing. One possible reason for the disagreement of the two expansions is CSW-like configurations which are not the product of two MHV amplitudes but are still non-zero residues of the link formula. In other words, we may have a pair of lines in the $C$-space with all negative helicity particles on one line. This is obviously not a Risager's term. For small $n$, as was said, the numerator does not allow such configurations in the expansion; in fact, the residue will be zero. But, for higher $n$ one can still have a valid expansion of residues, as the power counting shows, but there exist more terms which are not $\text{MHV}\times \text{MHV}$. These extra terms are in fact residues at infinity of the Risager's method. It would be interesting to explore these extra terms more precisely. 

\section*{Acknowledgments}
We would like to thank Freddy Cachazo for suggesting this problem, many helpful discussions and valuable comments on the draft. We also thank Song He for discussion on the supersymmetric residue calculations.~B.P. would like to thank AIMS-South Africa for the hospitality.~G.S. thanks Perimeter Institute for the hospitality during the early stages of this work. Research at Perimeter Institute is supported by the Government of Canada through Industry Canada and by the Province of Ontario through the Ministry of Research \& Innovation.~S.R. is supported in part by the NSERC of Canada and MEDT of Ontario.

\appendix

\section{Numerical Values for the 6- and 7-Graviton Residues and Risager Terms}
\label{app}
Here we present the numerical values for different channels given by the Risager diagrams and our CSW-like terms for a given set of external data for the 6- and 7-graviton cases. As described in section \ref{sec:regulator}, the residue computations are based on the use of a regulator $\epsilon$ for the minors of the integrand. We chose $\epsilon$ to take the value $0.0000001$. Also, note that the amplitude is the sum of Risager terms but is minus the sum of residues as is known from the residue theorem. So, here, signs of the values of the two columns are opposite. The biggest errors in the 6- and 7-graviton computations are $1.6\times 10^{-3}\%$ and $1.4\times 10^{-2}\%$ respectively.\\
\newline
Our data for the 6-graviton computation are as follows:

\begin{equation*}
\lambda_1= \begin{pmatrix}[r]
1 \\ 0
\end{pmatrix}, \lambda_2= \begin{pmatrix}
0 \\ 1
\end{pmatrix}, \lambda_3= \begin{pmatrix}
8 \\ 5
\end{pmatrix}, \lambda_4= \begin{pmatrix}
10 \\ 7
\end{pmatrix}, \lambda_5= \begin{pmatrix}[r]
-7 \\ 9
\end{pmatrix}, \lambda_6= \begin{pmatrix}
9 \\ 5
\end{pmatrix},
\end{equation*}
\begin{equation*}
\tilde\lambda_1= \begin{pmatrix}[r]
-50 \\ 35
\end{pmatrix}, \tilde\lambda_2= \begin{pmatrix}[r]
71 \\ -25
\end{pmatrix}, \tilde\lambda_3= \begin{pmatrix}[r]
-5 \\ 8
\end{pmatrix}, \tilde\lambda_4= \begin{pmatrix}
-4 \\ -8
\end{pmatrix}, \tilde\lambda_5= \begin{pmatrix}[r]
-7 \\ 4
\end{pmatrix}, \tilde\lambda_6= \begin{pmatrix}
9 \\ 1
\end{pmatrix},
\end{equation*}
and for the 7-graviton case we used:
\begin{equation*}
\lambda_1= \begin{pmatrix}
1 \\ 0
\end{pmatrix}, \lambda_2= \begin{pmatrix}
0 \\ 1
\end{pmatrix}, \lambda_3= \begin{pmatrix}
-9 \\ -5
\end{pmatrix}, \lambda_4= \begin{pmatrix}
-3 \\ -3
\end{pmatrix}, \lambda_5= \begin{pmatrix}[r]
9 \\ -6
\end{pmatrix}, \lambda_6= \begin{pmatrix}
-9 \\ -4
\end{pmatrix}, \lambda_7= \begin{pmatrix}
-10 \\ -8
\end{pmatrix},
\end{equation*}
\begin{equation*}
\tilde\lambda_1= \begin{pmatrix}
127 \\ 67
\end{pmatrix}, \tilde\lambda_2= \begin{pmatrix}
56 \\ 57
\end{pmatrix}, \tilde\lambda_3= \begin{pmatrix}[r]
-2 \\ 5
\end{pmatrix}, \tilde\lambda_4= \begin{pmatrix}
-2 \\ -2
\end{pmatrix}, \tilde\lambda_5= \begin{pmatrix}
-2 \\ -1
\end{pmatrix}, \tilde\lambda_6= \begin{pmatrix}[r]
7 \\ -9
\end{pmatrix}, \tilde\lambda_7= \begin{pmatrix}
7 \\ 10
\end{pmatrix}.
\end{equation*}

\pagebreak

\begin{table}[h]
\centering
\begin{tabular}{|l|d|d|l|}
 \hline
\multicolumn{1}{|c|}{Channel} & \multicolumn{1}{c|}{Risager's Term} & \multicolumn{1}{c|}{Residue} & \multicolumn{1}{c|}{Error} \\
  \hline
\{1,2\}\{3,4,5,6\} & \multicolumn{1}{c|}{$-3.8035$} & \multicolumn{1}{c|}{$3.8048$} & $3.4\times 10^{-4}\%$\\
\{1,3\}\{2,4,5,6\} & \multicolumn{1}{c|}{$0.027552142$} & \multicolumn{1}{c|}{$-0.027552138$} & $1.5\times 10^{-7}\%$\\
\{1,4\}\{2,3,5,6\} & \multicolumn{1}{c|}{$-0.00024281392$} & \multicolumn{1}{c|}{$0.00024281389$} & $1.2\times 10^{-7}\%$\\
\{1,5\}\{2,3,4,6\} & \multicolumn{1}{c|}{$-12.778$} & \multicolumn{1}{c|}{$12.757$} & $1.6\times 10^{-3}\%$\\
\{1,6\}\{2,3,4,5\} & \multicolumn{1}{c|}{$0.0099509309$} & \multicolumn{1}{c|}{$-0.0099509312$} & $3.0\times 10^{-8}\%$\\
\{2,3\}\{1,4,5,6\} & \multicolumn{1}{c|}{$-0.0014393998$} & \multicolumn{1}{c|}{$0.0014393997$} & $6.9\times 10^{-8}\%$\\
\{2,4\}\{1,3,5,6\} & \multicolumn{1}{c|}{$0.0001347570$} & \multicolumn{1}{c|}{$-0.0001347586$} & $1.2\times 10^{-5}\%$\\
\{2,5\}\{1,3,4,6\} & \multicolumn{1}{c|}{$0.2376382$} & \multicolumn{1}{c|}{$-0.2376353$} & $1.2\times 10^{-5}\%$\\
\{2,6\}\{1,3,4,5\} & \multicolumn{1}{c|}{$-0.0000115706946$} & \multicolumn{1}{c|}{$0.0000115706940$} & $5.2\times 10^{-8}\% $\\
\{3,4\}\{1,2,5,6\} & \multicolumn{1}{c|}{$-0.000011926600$} & \multicolumn{1}{c|}{$0.000011926618$} & $1.5\times 10^{-6}\% $\\
\{3,5\}\{1,2,4,6\} & \multicolumn{1}{c|}{$-0.0135260863$} & \multicolumn{1}{c|}{$0.0135260869$} & $4.4\times 10^{-8} \%$\\
\{3,6\}\{1,2,4,5\} & \multicolumn{1}{c|}{$-0.0010910475$} & \multicolumn{1}{c|}{$0.0010910470$} & $4.6\times 10^{-7}\% $\\
\{1,2,4\}\{3,5,6\} & \multicolumn{1}{c|}{$-0.00160980$} & \multicolumn{1}{c|}{$0.00160979$} & $6.2\times 10^{-6} \%$\\
\{1,2,5\}\{3,4,6\} & \multicolumn{1}{c|}{$15.50$} & \multicolumn{1}{c|}{$-15.48$} & $1.3\times 10^{-3}\%$\\
\{1,2,6\}\{3,4,5\} & \multicolumn{1}{c|}{$-0.0011519435$} & \multicolumn{1}{c|}{$0.0011519433$} & $1.7\times 10^{-7}\% $\\
\{1,3,4\}\{2,5,6\} & \multicolumn{1}{c|}{$-0.023460139$} & \multicolumn{1}{c|}{$0.023460140$} & $4.3\times 10^{-8} \% $\\
\{1,3,5\}\{2,4,6\} & \multicolumn{1}{c|}{$8.31380501\times  10^{-7}$} & \multicolumn{1}{c|}{$-8.31380545 \times 10^{-7}$} & $5.3\times 10^{-8} \%$\\
\{1,3,6\}\{2,4,5\} & \multicolumn{1}{c|}{$-0.0000245346$} & \multicolumn{1}{c|}{$0.0000245368$} & $9.0\times 10^{-5} \% $\\
\{1,4,5\}\{2,3,6\} & \multicolumn{1}{c|}{$-0.0005562082$} & \multicolumn{1}{c|}{$0.0005562084$} & $3.6\times 10^{-7} \%$\\
\{1,4,6\}\{2,3,5\} & \multicolumn{1}{c|}{$0.0069317909$} & \multicolumn{1}{c|}{$-0.0069317930$} & $3.0\times 10^{-7} \%$\\
\{1,5,6\}\{2,3,4\} & \multicolumn{1}{c|}{$6.4118974\times 10^{-7}$} & \multicolumn{1}{c|}{$-6.4118965\times 10^{-7}$} & $1.4\times 10^{-7} \%$\\
\hline
\end{tabular}
\caption{Numerical Values for 6-Graviton Computation}
\label{tab:myfirsttable}
\end{table} 
\newpage 

\begin{table}[h]
\centering
\begin{tabular}{|l|d|d|l|}
 \hline
\multicolumn{1}{|c|}{Channel} & \multicolumn{1}{c|}{Risager's Term} & \multicolumn{1}{c|}{Residue} & \multicolumn{1}{c|}{Error} \\
  \hline
\{1,2\}\{3,4,5,6,7\} & \multicolumn{1}{c|}{$-1.4302\times 10^{-7}$} & \multicolumn{1}{c|}{$1.4327\times 10^{-7}$} & $1.7\times 10^{-3}\%$\\
\{1,3\}\{2,4,5,6,7\} & \multicolumn{1}{c|}{$-0.0012650646$} & \multicolumn{1}{c|}{$0.0012650621$} & $2.0\times 10^{-6}\%$\\
\{1,4\}\{2,3,5,6,7\} & \multicolumn{1}{c|}{$4.075\times 10^{-8}$} & \multicolumn{1}{c|}{$-4.099\times 10^{-8}$} & $6.1\times 10^{-3}\%$\\
\{1,5\}\{2,3,4,6,7\} & \multicolumn{1}{c|}{$0.000054264493$} & \multicolumn{1}{c|}{$-0.000054264483$} & $1.8\times 10^{-7}\%$\\
\{1,6\}\{2,3,4,5,7\} & \multicolumn{1}{c|}{$-0.0003451595$} & \multicolumn{1}{c|}{$0.0003451593$} & $5.8\times 10^{-7}\%$\\
\{1,7\}\{2,3,4,5,6\} & \multicolumn{1}{c|}{$-0.0011835542$} & \multicolumn{1}{c|}{$0.0011835558$} & $1.4\times 10^{-6}\%$\\
\{2,3\}\{1,4,5,6,7\} & \multicolumn{1}{c|}{$-0.0024191261$} & \multicolumn{1}{c|}{$0.0024191232$} & $1.2\times 10^{-6}\%$\\
\{2,4\}\{1,3,5,6,7\} & \multicolumn{1}{c|}{$2.869570\times 10^{-7}$} & \multicolumn{1}{c|}{$-2.869555\times 10^{-7}$} & $5.2\times 10^{-6}\%$\\
\{2,5\}\{1,3,4,6,7\} & \multicolumn{1}{c|}{$2.553\times 10^{-7}$} & \multicolumn{1}{c|}{$-2.589\times 10^{-7}$} & $1.4\times 10^{-2}\%$\\
\{2,6\}\{1,3,4,5,7\} & \multicolumn{1}{c|}{$-0.0001381087$} & \multicolumn{1}{c|}{$0.0001381085$} & $1.4\times 10^{-6}\%$\\
\{2,7\}\{1,3,4,5,6\} & \multicolumn{1}{c|}{$0.021812$} & \multicolumn{1}{c|}{$-0.021809$} & $1.3\times 10^{-4}\%$\\
\{3,4\}\{1,2,5,6,7\} & \multicolumn{1}{c|}{$-0.0000389804$} & \multicolumn{1}{c|}{$0.0000389807$} & $7.7\times 10^{-6}\%$\\
\{3,5\}\{1,2,4,6,7\} & \multicolumn{1}{c|}{$0.000045447$} & \multicolumn{1}{c|}{$-0.000045450$} & $6.6\times 10^{-5}\%$\\
\{3,6\}\{1,2,4,5,7\} & \multicolumn{1}{c|}{$-0.002108237$} &\multicolumn{1}{c|}{ $0.002108243$} & $2.8\times 10^{-6}\%$\\
\{3,7\}\{1,2,4,5,6\} & \multicolumn{1}{c|}{$-0.00326858$} & \multicolumn{1}{c|}{$0.00326854$} & $1.2\times 10^{-5}\%$\\
\{1,2,4\}\{3,5,6,7\} & \multicolumn{1}{c|}{$-9.51196\times 10^{-8}$} & \multicolumn{1}{r|}{$9.51141\times 10^{-8}$} & $5.8\times 10^{-5}\%$\\
\{1,2,5\}\{3,4,6,7\} & \multicolumn{1}{r|}{$-3.365311\times 10^{-8}$} & \multicolumn{1}{r|}{$3.365362\times 10^{-8}$} & $1.5\times 10^{-5}\%$\\
\{1,2,6\}\{3,4,5,7\} & \multicolumn{1}{r|}{$5.972950\times 10^{-8}$} & \multicolumn{1}{r|}{$-5.972944\times 10^{-8}$} & $1.0\times 10^{-6}\%$\\
\{1,2,7\}\{3,4,5,6\} & \multicolumn{1}{c|}{$-0.0204886$} & \multicolumn{1}{c|}{$0.0204857$} & $1.4\times 10^{-4}\%$\\
\{1,3,4\}\{2,5,6,7\} & \multicolumn{1}{c|}{$-0.00033009914133$} & \multicolumn{1}{c|}{$0.00033009914132$} & $3.0\times 10^{-11}\%$\\
\{1,3,5\}\{2,4,6,7\} & \multicolumn{1}{c|}{$-0.0010859253$} & \multicolumn{1}{c|}{$0.0010859242$} & $1.0\times 10^{-6}\%$\\
\{1,3,6\}\{2,4,5,7\} & \multicolumn{1}{c|}{$-0.0058003020$} & \multicolumn{1}{c|}{$0.0058003065$} & $7.8\times 10^{-7}\%$\\
\{1,3,7\}\{2,4,5,6\} & \multicolumn{1}{c|}{$0.000767141$} & \multicolumn{1}{c|}{$-0.000767139$} & $2.6\times 10^{-6}\%$\\
\{1,4,5\}\{2,3,6,7\} & \multicolumn{1}{r|}{$5.82339379\times 10^{-6}$} & \multicolumn{1}{r|}{$-5.82339358\times 10^{-6}$} & $3.6\times 10^{-8}\%$\\
\{1,4,6\}\{2,3,5,7\} & \multicolumn{1}{c|}{$-0.000048608115$} & \multicolumn{1}{c|}{$0.000048608121$} & $1.2\times 10^{-7}\%$\\
\{1,4,7\}\{2,3,5,6\} & \multicolumn{1}{c|}{$-0.0016051200$} & \multicolumn{1}{c|}{$0.0016051227$} & $1.7\times 10^{-6}\%$\\
\{1,5,6\}\{2,3,4,7\} & \multicolumn{1}{c|}{$0.0001486804$} & \multicolumn{1}{c|}{$-0.0001486801$} & $2.0\times 10^{-6}\%$\\
\{1,5,7\}\{2,3,4,6\} & \multicolumn{1}{c|}{$0.0023658846$} & \multicolumn{1}{c|}{$-0.0023658860$} & $5.9\times 10^{-7}\%$\\
\{1,6,7\}\{2,3,4,5\} & \multicolumn{1}{c|}{$0.0001890601$} & \multicolumn{1}{c|}{$-0.0001890600$} & $5.3\times 10^{-7}\%$\\
\{2,3,4\}\{1,5,6,7\} & \multicolumn{1}{c|}{$0.0045052257$} & \multicolumn{1}{c|}{$-0.0045052235$} & $4.9\times 10^{-7}\%$\\
\{2,3,5\}\{1,4,6,7\} & \multicolumn{1}{c|}{$-0.0000926669$} & \multicolumn{1}{c|}{$0.0000926696$} & $2.9\times 10^{-5}\%$\\
\{2,3,6\}\{1,4,5,7\} & \multicolumn{1}{c|}{$0.002235755$} & \multicolumn{1}{c|}{$-0.002235760$} & $2.2\times 10^{-6}\%$\\
\{2,3,7\}\{1,4,5,6\} & \multicolumn{1}{r|}{$4.24184246\times 10^{-6}$} & \multicolumn{1}{r|}{$-4.24184210\times 10^{-6}$} & $8.5\times 10^{-8}\%$\\
\{2,4,5\}\{1,3,6,7\} & \multicolumn{1}{r|}{$-4.0081097\times 10^{-9}$} & \multicolumn{1}{r|}{$4.0081081\times 10^{-9}$} & $4.0\times 10^{-7}\%$\\
\{2,4,6\}\{1,3,5,7\} & \multicolumn{1}{c|}{$-0.0007386578$} & \multicolumn{1}{c|}{$0.0007386558$} & $2.7\times 10^{-6}\%$\\
\{2,4,7\}\{1,3,5,6\} & \multicolumn{1}{c|}{$0.008541409$} & \multicolumn{1}{c|}{$-0.008541422$} & $1.5\times 10^{-6}\%$\\
\{2,5,6\}\{1,3,4,7\} & \multicolumn{1}{c|}{$0.0001888818$} & \multicolumn{1}{c|}{$-0.0001888815$} & $1.6\times 10^{-6}\%$\\
\{2,5,7\}\{1,3,4,6\} & \multicolumn{1}{c|}{$-0.0031819569$} & \multicolumn{1}{c|}{$0.0031819561$} & $2.5\times 10^{-7}\%$\\
\{2,6,7\}\{1,3,4,5\} & \multicolumn{1}{c|}{$-0.0005354038$} & \multicolumn{1}{c|}{$0.0005354037$} & $1.9\times 10^{-7}\%$\\
\{3,4,5\}\{1,2,6,7\} & \multicolumn{1}{c|}{$0.000044443578$} & \multicolumn{1}{c|}{$-0.000044443556$} & $5.0\times 10^{-7}\%$\\
\{3,4,6\}\{1,2,5,7\} & \multicolumn{1}{c|}{$0.002633$} & \multicolumn{1}{c|}{$-0.002671$} & $1.4\times 10^{-2}\%$\\
\{3,4,7\}\{1,2,5,6\} & \multicolumn{1}{c|}{$0.00395962$} & \multicolumn{1}{c|}{$-0.00395958$} & $1.0\times 10^{-5}\%$\\
\{3,5,6\}\{1,2,4,7\} & \multicolumn{1}{c|}{$-0.006673719$} & \multicolumn{1}{c|}{$0.00667372$} & $1.2\times 10^{-6}\%$\\
\{3,5,7\}\{1,2,4,6\} & \multicolumn{1}{r|}{$2.363790\times 10^{-7}$} & \multicolumn{1}{r|}{$-2.363757\times 10^{-7}$} & $1.4\times 10^{-5}\%$\\
\{3,6,7\}\{1,2,4,6\} & \multicolumn{1}{c|}{$-0.000026712880$} & \multicolumn{1}{c|}{$0.000026712871$} & $3.4\times 10^{-7}\%$\\
  \hline
\end{tabular}
\caption{Numerical Values for 7-Graviton Computation}
\label{tab:secondtable}
\end{table}
\FloatBarrier
\newpage

\bibliographystyle{utphys}
\bibliography{gravitycsw}

\end{document}